\documentclass[12pt]{article}

\usepackage{epsf}

\begin{document}

\def\Journal#1#2#3#4{{#1} {\bf #2}, #3 (#4)}

\def\NP{{\em Nucl. Phys.} }
\def\PR{{\em Phys. Rev.} }
\def\AP{{\em Ann. of Phys.} }
\def\ZP{{\em Z. Phys.} }

\def\rme{{\rm e}}
\def\rmi{{\rm i}}
\def\rmx{{\rm x}}

\newcommand{\einh}{\frac{1}{2}}
\newcommand{\dreih}{\frac{3}{2}}
\newcommand{\be}{\begin{equation}}
\newcommand{\ee}{\end{equation}}
\newcommand{\bea}{\begin{eqnarray}}
\newcommand{\eea}{\end{eqnarray}}

\newcommand{\aein}{\hspace*{.5cm}}

\newcommand{\bsl}[1]{#1 \! \! \! /}

\newcommand{\lnp}[1]{{\cal L}_{N N \pi}^{#1}}
\newcommand{\lne}{{\cal L}_{N N \eta}}
\newcommand{\lng}{{\cal L}_{N N \gamma}}
\newcommand{\lnpg}{{\cal L}_{N N \pi \gamma}}
\newcommand{\lppg}{{\cal L}_{\pi \pi \gamma}}

\newcommand{\nres}[1]{N (#1)}
\newcommand{\dres}[1]{\Delta (#1)}

\newcommand{\lrp}[2]{{\cal L}_{#1 N \pi}^{#2}}
\newcommand{\lre}[1]{{\cal L}_{#1 N \eta}}
\newcommand{\lrg}[1]{{\cal L}_{#1 N \gamma}}

\newcommand{\frp}[1]{f_{\nres {#1} N \pi}}
\newcommand{\frg}[1]{f_{\nres {#1} N \gamma}}

\newcommand{\lnnv}{{\cal L}_{N N V}}
\newcommand{\lpgv}{{\cal L}_{V \pi \gamma}}

\newcommand{\real}{{\rm Re}}
\newcommand{\imag}{{\rm Im}}

\pagestyle{plain}
\pagenumbering{arabic}

\begin{center}
  {\LARGE Electromagnetic Couplings of Nucleon Resonances
  \footnote {Work supported by GSI Darmstadt and BMBF}$^,$\footnote {This paper forms part of the dissertation of T. Feuster}}
  \\[5mm]
  T. Feuster\footnote {e-mail:feuster@theorie.physik.uni-giessen.de} and
  U. Mosel\\[3mm]
  {\em Institut f\"ur Theoretische Physik, Universit\"at Giessen\\
       D--35392 Giessen, Germany}\\[3mm]
  UGI-96-06\\
\end{center}

\vspace{1.5cm}

%
%

\section*{Abstract}
An effective Lagrangian calculation of pion photoproduction including all 
nucleon resonances up to $\sqrt s = 1.7$ GeV is presented. We compare our 
results to recent calculations and show the influence of different width
parametrizations and offshell cutoffs on the photoproduction multipoles. We
determine the electromagnetic couplings of the resonances from a new fit to 
the multipole data.\\[3mm]

{\it PACS}: 13.60.Le\\
{\it Keywords}: electromagnetic couplings; baryon resonances; photoproduction

\newpage

%
%

\section{Introduction}

\aein In recent years effective Lagrangian models have been used to calculate 
pion and eta photoproduction on the nucleon 
\cite{lee,david,gg,benn,sauer,zhang}.

\aein For pion production these models were mainly 
applied to energies below the second resonance region, including nucleon 
Born terms, vector mesons and the $\dres {1232}$ resonance. It was found 
that proper unitarization is neccessary to describe the pion photoproduction 
amplitudes in all channels with $l \le 1$. Unitarity was guaranteed by
explicit inclusion of $\pi N$ rescattering \cite{lee} or by using Watsons 
theorem in one way or another \cite{david}.

\aein In the case of eta production the electromagnetic couplings of the 
$\nres {1535}$ resonance were extracted from a unitary calculation of the 
$E_{0+}^p$ pion multipole and the total $\gamma N \rightarrow \eta N$ 
cross section. These calculations include only the Born terms, vector 
mesons and the $S_{11}$ resonances. Offshell s- and u-channel 
$\dres {1232}$ and $\nres {1520}$ contributions to the pion 
production are neglected \cite{sauer} since only the $S_{11}$ 
$\pi N$ scattering channel was calculated to fit the hadronic properties
of the resonances.

\aein Unfortunately there is no "combined" model that extends the unitary
calculations done in the $\dres {1232}$ region up to the second or even
third resonance region ($\sqrt{s} \le 1.7$ GeV) including all resonances
and calculating all multipoles.

\aein As a first step Garcilazo et al. \cite{gg} neglected all rescattering
effects in the pion photoproduction and still found a resonable agreement
in most multipole channels with $l \le 2$. In the $\dres {1232}$ region the
remaining discrepancies in the $E_{1+}^{3/2}$ channel can be fully 
explained by the rescattering. But in this approach width parametrizations 
and additional cutoffs introduce unwanted ambiguities. Furthermore the role 
of the offshell contributions of the spin-$\dreih$ resonances was not 
investigated systematically. 

\aein Since the pion photoproduction is the main reaction to extract the
electromagnetic couplings of the nucleon resonances \cite{pdg} a careful 
examination of the influence of all degrees of freedom present in an
effective Lagrangian approach is neccesary. 

\aein The aim of this paper is therefore to show how these additional 
degrees of freedom influence the extracted electromagnetic couplings of 
the nucleon resonances given by \cite{gg}. For a set of possible 
combinations of width parametrizations and cutoffs we fit the 
photoproduction multipole data either with fixed or varying spin-$\dreih$ 
offshell contributions. It will be demonstrated that the extracted 
couplings depend heavily on these offshell contributions and that this 
problem can only be resolved in a complete calculation. Unfortunately the 
offshell contributions are most important in non-resonant multipole 
channels for energies away from the resonance position so that it is 
probably not possible to study the influence of the spin-$\dreih$ 
resonances separately. Since a full calculation has only
been carried out in the $\dres {1232}$ energy region \cite{scholten2} 
our calculation can be viewed as the starting point for further 
investigations. Especially an estimate of the size of different effects 
can be made.

\aein As an example for this we show the contribution of the 
$\nres {1520}$ to the $E_{0+}^p$ pion photoproduction multipole and compare 
it to the known rescattering effects. It can be seen that both effects are 
of the same order of magnitude and that therefore the determination of 
the $\nres {1535}$ couplings via the $E_{0+}^p$ multipole is influenced by 
the presence of the $\nres {1520}$.

%
%

\section{Model Lagrangians for pion photoproduction}

\aein Starting points are the interaction lagrangians for the hadronic and
electromagnetic coupling of the contributing particles. In the following
$m, M, M_R$ are the pion, nucleon and nucleon resonance mass, respectively.

\aein For the Born terms and the spin-$\einh$ resonances we use 
pseudovector (PV) $\pi N N$ and pseudoscalar (PS) $\eta N N$ coupling.
For the vector mesons ($\rho$ and $\omega$) we use the Lagrangians given 
in \cite{david}.

\aein Spin-$\einh$ ($\nres {1440}$, $\nres {1535}$, $\dres {1620}$ 
and $\nres {1650}$) resonances \cite{zhang}:
\bea
\lrp {R_{1/2}}{PS} &=&
- g_\pi \bar R \: \rmi 
\Gamma
{\vec \tau} {\vec \pi} N + h.c.  ,
\label{lnp12PS} \\
\lrp {R_{1/2}}{PV} &=&
- \frac{g_\pi}{M_{R} \pm M} \bar R
\Gamma_{\mu}
{\vec \tau} (\partial^{\mu} {\vec \pi}) N + h.c.  , 
\label{lnp12PV} \\
\lrg {R_{1/2}} &=&
e \bar R (g^s_1 + g^v_1 \tau_3)
\frac{\Gamma_{\mu \nu}}{4 M} N F^{\mu \nu} + h.c.  ,
\label{lng12}
\eea
where $R$ is the resonance spinor. The upper sign in (\ref{lnp12PV}) holds 
for even parity, the lower sign for odd parity resonances. The operators 
$\Gamma$, $\Gamma_{\mu}$ and $\Gamma_{\mu \nu}$ are given by
\bea
\Gamma = 1, \qquad \Gamma_{\mu} = \gamma_{\mu}, \qquad 
        \Gamma_{\mu \nu} = \gamma_{5} \sigma_{\mu \nu},
\label{gam_odd} \\
\Gamma = \gamma_{5}, \qquad \Gamma_{\mu} = \gamma_{5} \gamma_{\mu}, \qquad 
        \Gamma_{\mu \nu} = \sigma_{\mu \nu},
\label{gam_even}
\eea
where (\ref{gam_odd}) and (\ref{gam_even}) correspond to resonances of
odd and even parities, respectively. The magnetic couplings
for proton and neutron targets are $g^p_1 = g^s_1 + g^v_1$ and 
$g^n_1 = g^s_1 - g^v_1$. $F^{\mu \nu}$ represents the electromagnetic field
tensor.

\aein Spin-$\dreih$ ($\dres {1232}$, $\nres {1520}$ and $\dres {1700}$)
resonances:
\bea
\lrp {R_{3/2}}{} &=&
\frac{f_\pi}{m} \bar R^{\alpha} \Theta_{\alpha \mu} (z_{\pi})
\left [ \begin{array}{c} 1 \\ \gamma_5 \end{array} \right ]
{\vec T} (\partial^{\mu} {\vec \pi}) N + h.c.  , 
\label{lnp32} \\
\lrg {R_{3/2}} &=&
\frac{\rmi e g_1}{2 M} \bar R^{\alpha} \Theta_{\alpha \mu} (z_1)
\gamma_{\nu} 
\left [ \begin{array}{c} \gamma_5 \\ 1 \end{array} \right ]
T_3 N F^{\nu \mu} + \\
& & - \frac{e g_2}{4 M^2} \bar R^{\alpha} \Theta_{\alpha \mu} (z_2)
\left [ \begin{array}{c} \gamma_5 \\ 1 \end{array} \right ]
T_3 (\partial_{\nu} N) F^{\nu \mu} + h.c.  , \\
\Theta_{\alpha \mu} (z) &=&
g_{\alpha \mu} - \frac{1}{2} (1 + 2 z) \gamma_{\alpha} \gamma_{\mu}  .
\label{lng32}
\eea

For the $\nres {1520}$ the proton and neutron couplings $g_1^{p,n}$ and 
$g_2^{p,n}$ are used. The operator $\Theta_{\alpha \mu} (z)$ describes the 
offshell admixture of spin-$\einh$ fields. Some attempts have been made to 
fix the parameters $z$ by examing the Rarita-Schwinger equations and the
transformation properties of the interaction Lagrangians \cite{peccei,nath}. 
Since most of the arguments presented there do not hold for composite
particles and not all problems of interacting spin-$\dreih$ fields could
be solved, we treat these parameters as free and try to determine them by 
fitting the pion photoproduction multipoles. For a more detailed discussion 
see \cite{david}.

\aein The couplings (\ref{lng12}) - (\ref{lng32}) can easily be compared to 
other choices \cite{lee,gg,zhang}. The $R N \gamma$ Lagrangians differ only
by normalizations factors like $(M_R + M)/(2 M)$ or by use of the "Sachs"
type couplings in the $R_{3/2} N \gamma$ vertices.

\aein The Lagrangian $\lnnv$ for the vector meson coupling is choosen in 
analogy to the $N N \gamma$ case \cite{david}:
\bea
\lng &=& - e \bar N \left \{ \frac{(1 + \tau_3)}{2} \gamma_{\mu} A^{\mu} -
(\kappa^s + \kappa^v \tau_3) \frac{\sigma_{\mu \nu}}{4 M}
F^{\mu \nu} \right \} N, \\
\lnnv &=& - g_{NNV} \bar N \left \{ \gamma_{\mu} V^{\mu} -
K_V \frac{\sigma_{\mu \nu}}{2 M} V^{\mu \nu} \right \} N,
\label{lnnv}
\eea
where $V_{\mu}$ is the $\rho$ or $\omega$ field and 
$V_{\mu \nu} = \partial_{\nu} V_{\mu} - \partial_{\mu} V_{\nu}$ is the
the vector meson field tensor.
Otherwise it is not possible to use the VMD values for $K_V$ which are 
derived from the anomalous magnetic moments $(\kappa^s + \kappa^v \tau_3)$ 
of the nucleon. For means of comparison we use the same vector meson 
couplings as in \cite{lee,gg}:
\bea
g_{\omega \pi \gamma} = 0.313, \qquad
g_{\rho^{0} \pi \gamma} = 0.131, \qquad
g_{\rho^{\pm} \pi \gamma} = 0.103, \nonumber \\
g_{NN \omega} = 3 g_{NN \rho} = 7.98, \qquad
K_{\omega} = -0.12, \qquad
K_{\rho} = 3.71.
\label{veccoup}
\eea
For completeness we also give the electromagnetic vertex of the vector 
mesons \cite{david}:
\be
\lpgv = e \frac{g_{V\pi\gamma}}{4 m} 
\varepsilon_{\mu \nu \lambda \sigma}
F^{\mu \nu} V^{\lambda \sigma} \pi.
\ee

\begin{figure}[ht]
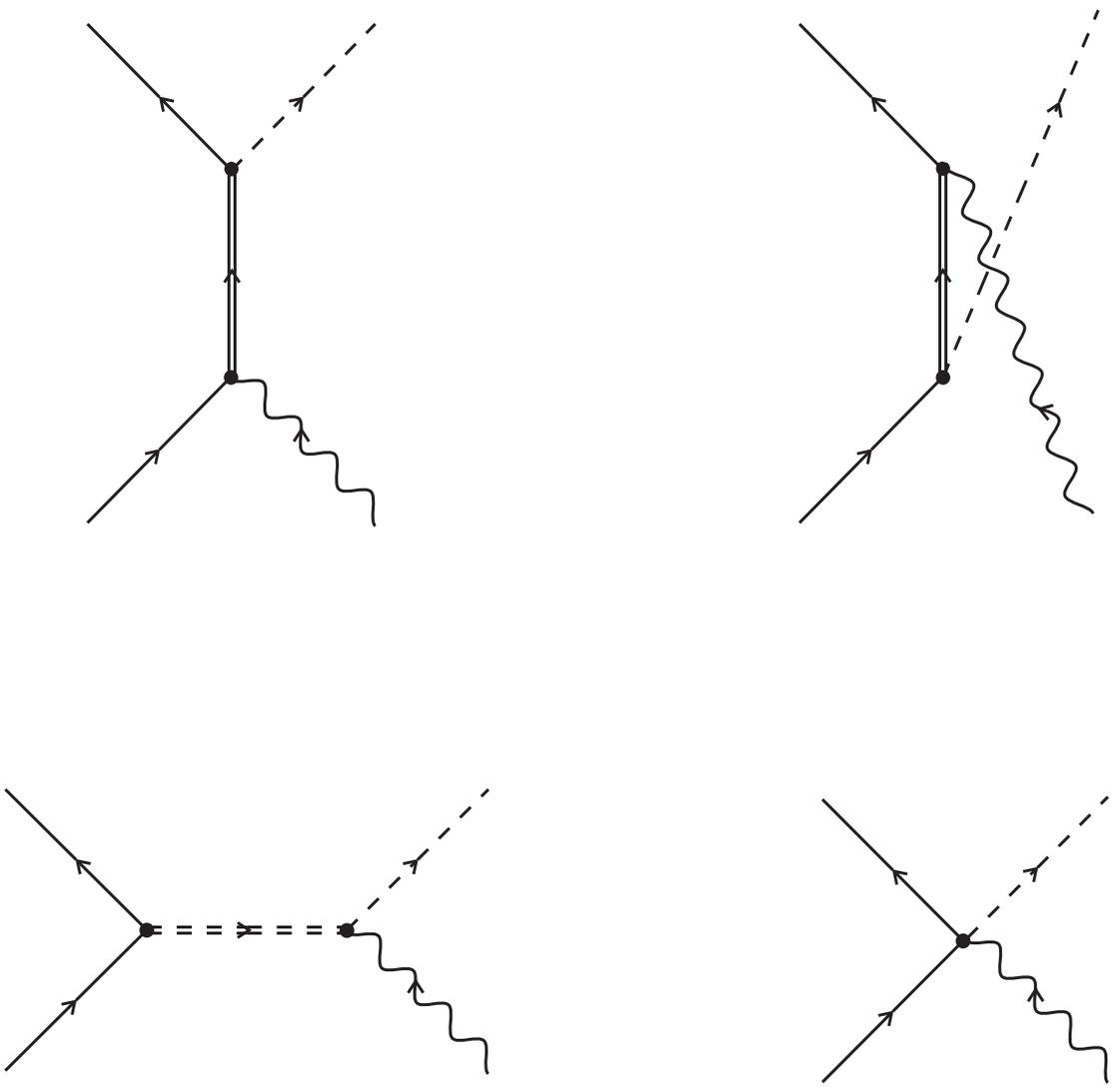

\caption{Feynman diagrams for pion photoproduction. From left to right:
direct graph, exchange graph, pion pole or vector meson graph, seagull graph.}
\label{diags}
\end{figure}

\aein In Fig. \ref{diags} we show the Feynman diagrams included in our
calculation. From the corresponding matrix elements 
we extract the photoproduction multipoles. In the first step the isospin
amplitudes are calculated from the physical amplitudes:
\bea
M^{3/2} &=& 
        M^{\pi^0} - \frac{1}{\sqrt{2}} M^{\pi^+} \nonumber \\
M^{1/2} &=& 
        M^{\pi^0} + \frac{1}{2 \sqrt{2}} M^{\pi^+} - \frac{3}{\sqrt{2}} M^{\pi^-}\nonumber \\
M^{0} &=& 
        \frac{1}{2 \sqrt{2}} \left ( M^{\pi^+} + M^{\pi^-} \right ) .
\eea
Projecting on total angular momentum $J$ we obtain the helicity amplitudes
\be
H^{I,J}_{\mu, \lambda}(W) =
\frac{e M}{8 \pi W} \: 2 
\int\limits^{\pi}_{0} d \theta \sin \theta \: 
M^{I}_{\mu, \lambda}(W, \theta) \: d^{J}_{\lambda, \mu}(\theta) .
\ee
Here $I$ denotes the isospin channel. $\mu$ and $\lambda$ are the final and 
inital helicities, respectively. The multipole amplitudes are now given by:
\bea
E^{I}_{0+}(W) &=& \frac{\sqrt{2}}{4} 
\left ( H^{I, \: 1/2}_{1/2, \: 1/2} - 
        H^{I, \: 1/2}_{1/2, \: -1/2} \right ) , \nonumber \\
M^{I}_{1-}(W) &=& \frac{- \sqrt{2}}{4} 
\left ( H^{I, \: 1/2}_{1/2, \: 1/2} + 
        H^{I, \: 1/2}_{1/2, \: -1/2} \right ) , \nonumber \\
E^{I}_{1+}(W) &=& \frac{\sqrt{2}}{8} 
\left \{ \frac{- 1}{\sqrt{3}}
\left ( H^{I, \: 3/2}_{1/2, \: 3/2} - 
        H^{I, \: 3/2}_{1/2, \: -3/2} \right )
        +
\left ( H^{I, \: 3/2}_{1/2, \: 1/2} - 
        H^{I, \: 3/2}_{1/2, \: -1/2} \right ) 
\right \} , \nonumber \\
M^{I}_{1+}(W) &=& \frac{\sqrt{2}}{8} 
\left \{ \sqrt{3}
\left ( H^{I, \: 3/2}_{1/2, \: 3/2} - 
        H^{I, \: 3/2}_{1/2, \: -3/2} \right )
        +
\left ( H^{I, \: 3/2}_{1/2, \: 1/2} - 
        H^{I, \: 3/2}_{1/2, \: -1/2} \right ) 
\right \} , \nonumber \\
E^{I}_{2-}(W) &=& \frac{\sqrt{2}}{8} 
\left \{ \sqrt{3}
\left ( H^{I, \: 3/2}_{1/2, \: 3/2} + 
        H^{I, \: 3/2}_{1/2, \: -3/2} \right )
        +
\left ( H^{I, \: 3/2}_{1/2, \: 1/2} + 
        H^{I, \: 3/2}_{1/2, \: -1/2} \right ) 
\right \} , \nonumber \\
M^{I}_{2-}(W) &=& \frac{\sqrt{2}}{8} 
\left \{ \frac{1}{\sqrt{3}}
\left ( H^{I, \: 3/2}_{1/2, \: 3/2} + 
        H^{I, \: 3/2}_{1/2, \: -3/2} \right )
        -
\left ( H^{I, \: 3/2}_{1/2, \: 1/2} + 
        H^{I, \: 3/2}_{1/2, \: -1/2} \right ) 
\right \} .
\eea
%

%
%

\section{Width parametrizations and cutoff functions}

\aein One of the ambiguities introduced in a tree level calculation of the 
photoproduction amplitudes comes from the inclusion of an energy dependent
decay width for the nucleon resonances. As Benmerrouche et al. \cite{muko}
have shown, unitarity is violated if the denominator of the resonance 
propagator is simply taken to be $(q^2 - M_R^2 + i M_R \Gamma(q^2))^{-1}$.
This problem can only be resolved in a K-Matrix approach to
meson-nucleon scattering and photoproduction as it has been done for the
$\dres {1232}$ region and the $E_{0+}$ multipole in \cite{david,sauer}. 

\aein Since we are mainly interested in the influence of different 
phenomenological width parametrizations we choose the free decay widths
in the pion, eta and two-pion channels calculated using the Lagrangians 
(\ref{lnp12PS}) - (\ref{lnp32}) times one of the following cutoff 
factors \cite{gg,moniz}:
\be
F^{G} (X) =
\frac{2}{1 + (X/X_0)^{\alpha_G}} , \quad
F^{M} (X) =
\left ( \frac{X_0 + x}{X + x} \right )^{\alpha_M} .
\label{cutoffs}
\ee
In the pion and eta decay channels $X = p^{2}$, where $p$ is the 
three-momentum of the outgoing meson, for the two-pion case $X$ is the "free" 
energy $X = \sqrt s - M - 2 m$. In our work we calculate the total width by 
summing up the possible partial decay widths using 
{\it the same cutoff parameters for all decay channels and nucleon 
resonances}. This was done in order to avoid introducing too many free 
parameters. The value $x = (0.3 {\rm GeV})^2$ ($x = 0.3$ GeV for 
two-pion decays) was fixed in $\pi N$ scattering \cite{moniz}. 

\aein In their model Garcilazo et al. \cite{gg} used a simple 
parametrization of the {\it total} decay width:
\be
\Gamma (s) = \Gamma_{0} (p/p_{0})^{2l + 1}
\frac{2}{1 + (p/p_0)^{2l + 3}},
\label{ggwidth}
\ee
with $\Gamma_{0}$, $p_{0}$ being the width and the pion momentum at
$s = M_{R}^{2}$. This description might be useful for the 
$\dres {1232}$ channel, but fails to reproduce the $\nres {1440}$ and 
$\nres {1535}$ widths since the latter resonances have strong two-pion and 
eta decay channels, respectively. As an improved description, showing the 
same threshold and high energy behaviour we used
\be
\Gamma (s) = \sum_{i = \pi, \eta, \pi\pi} \Gamma^{i}_{0} (s) F^{G} (X^{i})
\label{ourwidth}
\ee
for the width in each possible decay channel. $\Gamma^{i}_{0} (s)$ is
the free decay width. The difference of both descriptions for the 
$\dres {1232}$ width is less than 2\% around the resonance positions and of 
the order of 6\% close to threshold. The parametrization (\ref{ourwidth}) 
therefore allows to treat the decay channels consistently but retains the 
main features of the one used in \cite{gg}.

\aein For the cutoff exponents we use $\alpha_M = l + 1$ for pion and eta 
decays and $\alpha_M = 2$ for the two-pion decay
\footnote{In the case of (pseudo)vector $\pi,\eta N$ coupling of the 
spin-$\einh$ resonances one has an additional factor $(\sqrt s + m)^2$ in 
the free decay width. Therefore we choose $\alpha_M = l + 2$ in this case to 
have the same asymptotic behaviour for scalar and vector coupling.}. 
For $\alpha_G$ 
Garcilazo et al. used $(2 l + 3) / 2$ (\ref{ggwidth}), but we have chosen 
$\alpha_G = \alpha_M$ to have the same high energy behavior for both cutoff 
parametrizations.

\aein The free widths in the pion and eta channels were calculated from
the corresponding decay diagram for an "onshell" resonance with mass 
$M_R = \sqrt s$ \cite{gg}. 
The results for the widths differ by a factor $(\sqrt s + M_{R}) / 2 M_{R}$ 
from those found from calculating the $\pi N$ scattering K-Matrix with the 
given couplings \cite{david}. This factor stems from the projection 
into the proper partial wave channel that is necessary in the latter case.
It can easily be absorbed into the cutoff parametrizations. For the 
two-pion branch we choose 
$\Gamma_{N \pi \pi} = \Gamma_{N \pi \pi}^0 X / X_0$, with
$\Gamma_{N \pi \pi}^0$ being the free two-pion decay width. 
This parametrizes the three-particle phase space \cite{benn, chiang}.

\aein Upon including cutoff factors in the width parametrizations 
consistency would demand a factor $\sqrt {F^{G,M}}$ in the 
corresponding meson nucleon Lagrangians, that has been ignored so far in 
most of the calculations. We will show how the extracted couplings depend
on such an extra factor.

\aein Garcilazo et al. have shown that in their calcuation it is not 
possible to reproduce the measured pion photoproduction multipoles 
without introducing an additional factor 
$\Lambda_u^2 / (\Lambda_u^2 + p^2), \Lambda_u = 0.3$ GeV for 
the u-channel resonance diagrams. This is because of the high energy 
divergence of these contributions which are not reduced by the 
$(u - M_R^2)$ denominator in the propagator. Such a cutoff is at this 
stage purely phenomenological. Furthermore in eta photoproduction a similar 
cutoff $(\Lambda_V^2  - m_V^2) / (\Lambda_V^2 - t), \Lambda_V = 1.2$ GeV 
is used at the $V N N$ vertex \cite{zhang}. The dependence of the couplings 
on these cutoffs will also be considered.

%
%

\section{Photoproduction multipoles}

\begin{table}
\begin{center}
\renewcommand{\arraystretch}{1.2}
\begin{tabular}{|l|rrrr|rrrrr|}
\hline
 & \multicolumn{4}{|c|}{Garcilazo} & \multicolumn{5}{c|}{PDG 94} \\
 & $M_R$ & $\Gamma_{\pi}$ & $\Gamma_{\eta}$ & $\Gamma_{2\pi}$ & 
   $M_R$ & Pole & $\Gamma_{\pi}$ & $\Gamma_{\eta}$ & $\Gamma_{2\pi}$ \\
\hline\hline
$\dres {1232}$ & 1.215 & 0.106 & --- & --- & 1.232 & 1.210 & 0.120 &  ---  &  --- \\
$\nres {1440}$ & 1.430 & 0.140 & --- & --- & 1.440 & 1.370 & 0.228 &  ---  & 0.122 \\
$\nres {1520}$ & 1.505 & 0.070 & --- & --- & 1.520 & 1.510 & 0.066 &  ---  & 0.054 \\
$\nres {1535}$ & 1.500 & 0.060 & --- & --- & 1.535 & 1.500 & 0.068 & 0.064 & 0.018 \\
$\dres {1620}$ & 1.620 & 0.040 & --- & --- & 1.620 & 1.600 & 0.038 &  ---  & 0.112 \\
$\nres {1650}$ &  ---  &  ---  & --- & --- & 1.650 & 1.655 & 0.105 &  ---  & 0.019 \\     
$\dres {1700}$ & 1.700 & 0.050 & --- & --- & 1.700 & 1.660 & 0.045 &  ---  & 0.255 \\
\hline
\end{tabular}
\renewcommand{\arraystretch}{1.0}
\caption{Masses and widths used in this work (in GeV). The values of 
Garcilazo et al. were used to check our calculations (they 
did not include the \protect$\nres {1650}$ and gave no values for the 
eta and two-pion decay branches). For the PDG case {\it Pole} gives the 
avarage values for the real part of the pole positions.}
\label{particles}
\end{center}
\end{table}

\aein In Table \ref{particles} we show the values for masses and widths of
the nucleon resonances used in this work. The first set is that of 
Garcilazo et al.. while the second contains the mean values given by the 
Particle Data Group \cite{pdg}. Throughout this work the latter values were 
used. We then extracted the hadronic couplings of the resonances using the
formula for the free decay in the corresponding channels. The sign of the
couplings was choosen to be positive. The electromagnetic 
couplings were determined in fits to the photoproduction multipoles 
using either the resonance masses or the pole positions from \cite{pdg}
\footnote{To check the influence of the $\rho NN$ tensor coupling we
performed one fit (\# 5) using $K_{V}$ = 6.6. With this choice we find a 
$\chi^2$ value that is 2\% lower than the one given in Table \ref{chi2}.
The final parameter estimates show only small deviations from the values
found with $K_{V}$ = 3.71.}.

\aein As a check of our code we compared our results to the figures given by 
Garcilazo et al. \cite{gg}. While we use the same amplitudes as in eqn. (42) 
of ref. \cite{gg}, we are unable to reproduce their results. Only after 
changing the sign of $K_{V}$ in (\ref{lnnv}) we found exact agreement with 
the calculated amplitudes there. As a further check we calculated the 
total cross section both from the multipoles and directly using the 
Feynman matrix elements for the different diagrams. In order to check 
our $R N \gamma$ couplings we reproduced the calculation of Pascalutsa and 
Scholten \cite{scholten} for Compton scattering on the proton.

\aein Since we want to compare the influence of the different 
parametrizations in Table \ref{chi2} we only give
$\chi^2_{norm} = \chi^2/\chi^2_{GG}$ with $\chi^2_{GG}$ being the value 
extracted from the corrected model of Garcilazo et al.. Because of the 
disagreement in the resulting amplitudes mentioned above we first 
calculated $\chi^2_{GG}$ within the model given in \cite{gg} but using our 
vector meson couplings from (\ref{veccoup}). This leads to an increase 
of $\chi^2_{GG}$ of 20\% as compared to the original calculation of 
Garcilazo et al.. We then refitted the couplings to obtain a better 
$\chi^2_{norm}$ of  $\approx 0.49$ (Fit 1 in Table \ref{chi2}).

\aein In \cite{gg} the electromagnetic couplings were extracted from the 
experimental values for $A_{1/2, 3/2}$ on the resonance points and the
$z_i$ parameters of the spin-$\dreih$ resonances were fixed to the values 
given by Peccei \cite{peccei}, $z_i = 0.25$. In order to reproduce the 
multipole data the masses and widths of the resonances were then adjusted. 
From comparing the fits 1 and 2 in Table \ref{chi2} it is clear 
that the quality of the fit depends strongly on the values of the resonance 
masses and widths used in \cite{gg}. Since the extraction of the 
$A_{1/2, 3/2}$ parameters depends on a model for pion photoproduction, we 
have chosen to determine the electromagnetic couplings 
by fitting the experimental multipoles over the whole energy range 
instead of the resonance masses and widths, which we take from \cite{pdg}.

\aein The final $\chi^2$ values are given in Table \ref{chi2} for the use of 
both the resonance masses and poles. In general one can see that better fits 
can be found by using the pole positions instead of the resonance masses. 
The masses are normally determined in K-Matrix calculations of $\pi, \eta N$ 
scattering, whereas the the pole positions are taken directly from the 
corresponding experimental multipoles.

\begin{table}[t]
\begin{center}
\renewcommand{\arraystretch}{1.2}
\begin{tabular}{|r|l|l|l|l|l|l|rr|}
\hline
 & Resonance & Width  & $R N \pi$ & u channel & $V N N$ & offshell      & \multicolumn{2}{c|}{$\chi^2_{norm}$} \\
 & values    & cutoff & cutoff    & cutoff    & cutoff  & parameter $z$ & Masses & Poles \\
\hline\hline
 1 & Garcilazo & $F^G$  & no        & yes       & no      & z = -0.25     & \multicolumn{2}{c|}{$0.49^1$} \\
 2 & PDG 94    & $F^G$  & no        & yes       & yes     & z = -0.25     & 1.95 & 1.47 \\
 3 & PDG 94    & $F^G$  & no        & yes       & yes     & ${\rm Davidson}^2$ & 6.37 & 6.01 \\
 4 & PDG 94    & $F^G$  & no        & yes       & yes     & ${\rm Davidson}^3$ & 2.88 & 2.13 \\
 5 & PDG 94    & $F^G$  & no        & yes       & yes     & Fit           & 1.47 & 0.51 \\
 6 & PDG 94    & $F^M$  & no        & yes       & yes     & Fit           & 1.55 & 0.66 \\
 7 & PDG 94    & $F^G$  & yes       & yes       & yes     & Fit           & 2.66 & 1.93 \\
 8 & PDG 94    & $F^M$  & yes       & yes       & yes     & Fit           & 2.33 & 1.33 \\
 9 & PDG 94    & $F^G$  & no        & yes       & no      & Fit           & 1.51 & 0.47 \\
10 & PDG 94    & $F^G$  & no        & no        & yes     & Fit           & 1.71 & 1.73 \\
\hline
\end{tabular}
\renewcommand{\arraystretch}{1.0}
\caption{$\chi^2_{norm}$ of our fits as described in the text, using the 
different width parametrizations and cutoff factors. The first values give the 
$\chi^2_{norm}$ for the calculations with the resonance mass, the second are for 
the pole positions.
($^1$ : Refitted couplings within the model of Garcilazo et al.. The first set 
of Table \protect\ref{particles} was used, but the vector meson couplings 
from (\protect\ref{veccoup}), 
$^2$ : The values from 
\protect\cite{david} were used for all spin-$\dreih$ resonances, 
$^3$ : The offshell parameters of the $\nres {1520}$ and $\dres {1700}$ 
were allowed to vary)}
\label{chi2}
\end{center}
\end{table}

\aein In total 10 fits were performed for different combinations of
resonance data and cutoff factors. This allows to investigate the 
sensitivity of the extracted resonance parameters on the various 
parametrizations given in the last section. Table \ref{chi2} shows that 
the quality of all fits with the exception of no. 3 is comparable, with
fits 5, 6 and 9 being the best. The equal quality of fits 5 and 6 shows
that the shape of the width cutoffs (\ref{cutoffs}) is not essential.

\aein In one case (fit 3 in Table \ref{chi2}) the offshell parameters 
$z_i$ of the spin-$\dreih$ resonances were fixed to the values given by 
\cite{david}, $z_{\pi} = -0.24, z_1 = 2.39, z_2 = -0.53$. With this choice 
the overall $\chi^2$ increases by a factor of about 9 (see Fig. \ref{e0pp}). 
As is shown in Fig. \ref{e0pp2} for the $E_{0+}^{p}$ multipole this is due 
to the offshell contribution of the $D_{13}$ $\nres {1520}$ 
resonance which depends strongly on the choice 
of the $z_i$ parameters. To investigate this dependence in more detail we 
have performed one fit (4) where only the $\dres {1232}$ offshell 
parameters were taken from \cite{david} and the values for the $\nres {1520}$ 
and $\dres {1700}$ were allowed to vary. The resulting lower $\chi^2$ shows 
that the multipole data are highly sensitive to the $\nres {1520}$ offshell 
contributions. Especially the $E_{0+}^{p}$ multipole imposes strict limits 
on the $z_i$ parameters. 

\aein When the $\dres {1232}$ offshell parameters are also fitted to the data,
$\chi^2$ decreases for both choices of the cutoff factor $F^{G,M}$. The final
values for the $z_i$'s differ strongly from those given by Davidson et al..
This is probably due to the missing rescattering in our calculation.
Both offshell contributions and rescattering effects are most effective
in channels that are not strongly dominated by one resonance 
(eg. in $E_{1+}^{3/2}$). So during the fitting the $z_i$ parameters adjust to 
compensate for the lack of rescattering even though both effects result from
totally different physical mechanisms. 

\aein As shown in Table \ref{chi2} two fits (7 and 8) were made using a 
cutoff factor $\sqrt {F^{G,M}}$ at the $R N \pi$ vertices. Both of these 
cutoffs show the same high energy dependence (see Eq. (\ref{cutoffs})). Below 
threshold they are larger than 1 and thus enhance the resonance contributions 
(about a factor $\sqrt 2$ for $F^G$ and $2 - 30$ for $F^M$ depending on the 
mass and the angular momentum of the decay pion). For higher energies however, 
they both lead to about the same reduction of the corresponding amplitudes. 
As Table \ref{parms} shows, which compares one of those fits (7) with the
fit no. 5, the extracted electromagnetic couplings are drastically different 
in both cases. This was to be expected since both lead to a rather large 
$\chi^2$ value and therefore do not determine the resonance parameters 
very accurately. When using the resonance pole positions the final 
couplings for the fits 5 and 7 show a somewhat better agreement 
(Table \ref{parmsp}). 

\aein When using the resonance pole positions part of the rescattering 
effects are taken into account by the mass shift. This is the main reason for
the lower $\chi^2$ values found in these fits. Already the shift of the 
$\dres {1232}$ resonance can explain a large part of this effect. Since the
corresponding $M_{1+}^{3/2}$ multipole is dominant in the pion 
photoproduction, the value of the $\dres {1232}$ mass enters crucially into 
the calculations. As can be seen from Table \ref{particles} this value
is nearly the same in our calculation and in the work of Garcilazo et al..
So it was to be expected that their model leads to about the same
$\chi^2$ values as our best fits using the resonance pole positions.

\aein A comparison of the final parameter estimates in Table \ref{parmsp} shows
that the $\dres {1323}$ couplings are rather insensitive to the different 
prescriptions used in the fits. The offshell parameters of Fits 5 and 9 are
comparable while the result of Fit 7 shows large deviations. Since there is no 
interference with other resonance contributions in the $\dres {1323}$ energy 
region these differences are directly related to the different cutoffs used in 
the fits. For the $\nres {1520}$ and $\dres {1700}$ resonances this is not 
true. In this energy region we have resonances in most of the channels where 
there are also large offshell contributions. Therefore the $z_{i}$ parameters are 
not as uniquely determined as in the $\dres {1323}$ case. By comparing the 
Fits 5 and 7 with 9 one sees that different combinations of resonance couplings 
of the spin-$\einh$ resonances and offshell parameters can lead to about the 
same $\chi^2$. Especially for resonances with weak electromagnetic couplings 
($\dres {1620}$ and $\nres {1650}$) this leads to drastically different parameters.

\aein In order to check the dependence of the couplings on the use of the
u-channel and vector meson cutoffs we tried to fit the data without either
of these. The corresponding $\chi^2$ values are given in the last two lines 
of Table \ref{chi2}. Comparing fits 5 and 9 one can see that these are not 
very sensitive to a cutoff at the VNN vertex. Only for the pole positions we 
find differences to the fits done with this cutoff. The lower $\chi^2$ value 
results from cancellations of different contributions at higher energies. 
These can only take place if the vector meson amplitude is not reduced by 
a cutoff. In Tables \ref{parms} and \ref{parmsp} we give the extracted 
couplings for fit 9.

\begin{table}
\begin{center}
\renewcommand{\arraystretch}{1.2}
\begin{tabular}{|l|rr||l|rrrrr||c|}
\hline
 \multicolumn{3}{|c||}{Spin - $\einh$} & 
 \multicolumn{6}{c||}{Spin - $\dreih$} & \\
\multicolumn{2}{|r}{$g_p$} & $g_n$ &
   \multicolumn{2}{|r}{$g_1$} & $g_2$ & $z_{\pi}$ & $z_1$ & $z_2$ & \raisebox{1.5ex}[-1.5ex]{Fit} \\
\hline\hline
$\nres {1440}$ & -0.448 &  0.115 & 
        $\dres {1232}$ &  4.774 &  8.467 & -0.308 & -1.141 &  0.666 & \\
$\nres {1535}$ &  0.797 & -0.413 & 
                &  3.295 &  3.020 &        &        &        & \\
$\dres {1620}$ & -0.348 &   ---  & 
        \raisebox{1.5ex}[-1.5ex]{$\nres {1520}$} & -1.170 &  1.110 & 
                \raisebox{1.5ex}[-1.5ex]{-2.492} & 
                \raisebox{1.5ex}[-1.5ex]{-0.141} & 
                \raisebox{1.5ex}[-1.5ex]{-0.103} & 
 \raisebox{1.5ex}[-1.5ex]{\Large 5} \\
$\nres {1650}$ &  0.092 &  0.185 & 
        $\dres {1700}$ &  1.417 &  2.885 &  0.098 & -0.460 &  0.801 & \\
\hline\hline
$\nres {1440}$ & -0.404 &  0.313 & 
        $\dres {1232}$ &  5.478 &  7.611 & -0.594 &  0.050 &  1.499 & \\
$\nres {1535}$ &  0.649 & -0.544 & 
        &  4.914 &  5.061 &        &        &        & \\
$\dres {1620}$ & -0.131 &   ---  & 
        \raisebox{1.5ex}[-1.5ex]{$\nres {1520}$} & -1.387 &  0.677 & 
                \raisebox{1.5ex}[-1.5ex]{-0.617} &  
                \raisebox{1.5ex}[-1.5ex]{0.126} &  
                \raisebox{1.5ex}[-1.5ex]{0.889} & 
 \raisebox{1.5ex}[-1.5ex]{\Large 7} \\
$\nres {1650}$ &  0.154 & -0.023 & 
        $\dres {1700}$ &  1.038 &  2.542 &  0.170 &  0.131 & -2.292 & \\
\hline\hline
$\nres {1440}$ & -0.442 &  0.101 & 
        $\dres {1232}$ &  4.784 &  5.179 & -2.402 & -0.300 & -0.136 & \\
$\nres {1535}$ &  0.692 & -0.209 & 
        &  3.095 &  3.003 &        &        &        & \\
$\dres {1620}$ & -0.109 &   ---  & 
        \raisebox{1.5ex}[-1.5ex]{$\nres {1520}$} & -1.839 &  0.018 & 
                \raisebox{1.5ex}[-1.5ex]{-1.817} &  
                \raisebox{1.5ex}[-1.5ex]{-0.092} &  
                \raisebox{1.5ex}[-1.5ex]{-0.024} & 
 \raisebox{1.5ex}[-1.5ex]{\Large 9} \\
$\nres {1650}$ &  0.089 &  0.186 & 
        $\dres {1700}$ &  1.625 &  3.242 & -1.405 & -0.266 &  0.042 & \\
\hline
\end{tabular}
\renewcommand{\arraystretch}{1.0}
\caption{Final parameter estimates for fits 5, 7 and 9 of Table 
\protect\ref{chi2} using the resonance masses. 
Left side: spin-$\einh$, right side: spin-$\dreih$
resonances (For the $\nres {1520}$ proton couplings are given in the first,
neutron couplings in the second line).}
\label{parms}
\end{center}
\end{table}

\begin{table}
\begin{center}
\renewcommand{\arraystretch}{1.2}
\begin{tabular}{|l|rr||l|rrrrr||c|}
\hline
 \multicolumn{3}{|c||}{Spin - $\einh$} & 
 \multicolumn{6}{c||}{Spin - $\dreih$} & \\
\multicolumn{2}{|r}{$g_p$} & $g_n$ &
   \multicolumn{2}{|r}{$g_1$} & $g_2$ & $z_{\pi}$ & $z_1$ & $z_2$ & \raisebox{1.5ex}[-1.5ex]{Fit} \\
\hline\hline
$\nres {1440}$ & -0.400 &  0.110 & 
        $\dres {1232}$ &  5.416 &  6.612 &  1.400 & -0.293 & -0.394 & \\
$\nres {1535}$ &  0.623 & -0.583 & 
                &  3.449 &  3.003 &        &        &        & \\
$\dres {1620}$ & -0.144 &   ---  & 
        \raisebox{1.5ex}[-1.5ex]{$\nres {1520}$} & -0.307 &  1.862 & 
                \raisebox{1.5ex}[-1.5ex]{-2.418} & 
                \raisebox{1.5ex}[-1.5ex]{-0.158} & 
                \raisebox{1.5ex}[-1.5ex]{-0.160} & 
 \raisebox{1.5ex}[-1.5ex]{\Large 5} \\
$\nres {1650}$ &  0.205 &  0.200 & 
        $\dres {1700}$ &  1.895 &  3.921 & -0.606 & -1.120 &  0.124 & \\
\hline\hline
$\nres {1440}$ & -0.403 &  0.137 & 
        $\dres {1232}$ &  5.498 &  6.220 & -0.474 & -0.196 &  1.212 & \\
$\nres {1535}$ &  0.733 & -0.516 & 
        &  4.899 &  5.074 &        &        &        & \\
$\dres {1620}$ & -0.003 &   ---  & 
        \raisebox{1.5ex}[-1.5ex]{$\nres {1520}$} & -0.378 &  0.719 & 
                \raisebox{1.5ex}[-1.5ex]{-2.326} & 
                \raisebox{1.5ex}[-1.5ex]{-0.047} &  
                \raisebox{1.5ex}[-1.5ex]{0.148} & 
 \raisebox{1.5ex}[-1.5ex]{\Large 7} \\
$\nres {1650}$ &  0.279 &  0.147 & 
        $\dres {1700}$ &  1.335 &  2.694 & -0.398 & -1.032 &  0.204 & \\
\hline\hline
$\nres {1440}$ & -0.401 &  0.186 & 
        $\dres {1232}$ &  5.099 &  5.836 &  0.949 & -0.310 & -0.483 & \\
$\nres {1535}$ &  0.627 & -0.474 & 
        &  3.004 &  3.047 &        &        &        & \\
$\dres {1620}$ & -0.005 &   ---  & 
        \raisebox{1.5ex}[-1.5ex]{$\nres {1520}$} & -0.068 &  1.265 & 
                \raisebox{1.5ex}[-1.5ex]{-0.123} & 
                \raisebox{1.5ex}[-1.5ex]{1.390} &  
                \raisebox{1.5ex}[-1.5ex]{ 0.267} & 
 \raisebox{1.5ex}[-1.5ex]{\Large 9} \\
$\nres {1650}$ &  0.045 &  0.011 & 
        $\dres {1700}$ &  1.516 &  2.176 &  0.016 &  0.978 & -2.297 & \\
\hline
\end{tabular}
\renewcommand{\arraystretch}{1.0}
\caption{Same as Table \protect\ref{parms}, but for the resonance pole
positions.}
\label{parmsp}
\end{center}
\end{table}

\aein For a u-channel cutoff the situation is different. 
Gracilazo et al. showed that they needed this cutoff to supress the
contribution to the multipoles from the crossed resonance diagrams. A 
detailed analysis of their results shows that the $\dres {1232}$ 
accounts for most of the divergence they find. Our fit indicates that this 
conclusion strongly depends on the choice of the offshell parameters 
$z_i$ used. If these are allowed to vary, we find parameter sets that 
lead to large cancellations between the offshell contributions of 
different resonances. However, one can see from the fit using the resonance 
poles that the $\chi^2$ is not reduced in this case as it was for the other 
fits. This shows that the additional degrees of freedom from the $z_i$ 
parameters are used to mock up the effect of an u-channel cutoff without
leading to an improvement of other features. We thus conclude, in agreement 
with \cite{gg}, that without such an u-channel cutoff no satisfactory 
description of the multipole data can be found.

\aein In addition to the $E_{0+}^{p}$ multipole we show in figs. \ref{ipall} -
\ref{inall} the result for all multipoles calculated using the parameters of
Fit 5 using the resonance pole positions. In general we have good agreement with 
the experimental data in all channels.

\aein Clearly visible is the strong peak in the $E_{0+}^{p}$ multipole because 
of the reduced $\nres {1535}$ mass. In the $M_{1\pm}^{p}$ channels 
the $\nres {1520}$ offshell contributions are most prominent. Especially for the 
imaginary part this stands in contrast to the experimental values. For the 
$z_{i}$ parameters used by Garcilzao et al. only the $M_{1+}^{p}$ multipole 
shows this behaviour. But no choice of the offshell parameters could be found 
that reduces the effect in both multipole channels. It would be intereseting to 
see the effect of rescattering on this results since on the Lagrangian level only 
a drastic cutoff of the crossed resonance diagrams could remove this contributions.

\aein The missing rescattering is obvious in the $E_{1+}^{3/2}$ channel. Here it 
is well know that only a complete K-matrix calculation can reproduce the data 
\cite{muko}. The same seems to be true for the $M_{2-}^{n}$ multipole where one 
has as similar situation. An otherwise strong spin-$\dreih$ resonance 
($\nres {1520}$) with a weak coupling into this particular channel. Therefore 
this multipole would also be very sensitive to rescattering effects.

%
%

\section{Conclusions}

\aein In this work we have determined the electromagnetic couplings of
nucleon resonances by fitting the pion photoproduction multipoles for 
various choices of resonance values and cutoffs. The importance of 
pion photoproduction as the main reaction to extract these couplings 
\cite{pdg} shows how neccessary a detailed understanding of all 
model dependencies is. It was found that different values of the 
electromagnetic couplings in combination with different choices for the 
width parametrizations lead to fits of equal quality. Only for the 
resonances with strong electromagnetic decay channels the final coupling 
parameters are comparable. For the other resonances 
($\dres {1620}$ and $\nres {1650}$) the couplings are not well determined by 
this calculation.

\aein For the use of the resonance masses all extracted couplings are very 
sensitive to the cutoffs used in the calculation. We find resonably stable 
electromagnetic couplings only for the fits with the resonance pole 
positions. The values that we then find are in agreement with those given 
by other calculations \cite{pdg}. 

\aein The other important point in our view is the investigation of the
offshell contributions of the spin-$\dreih$ resonances. Besides the 
resonance mass values the offshell parameters $z_i$ have the biggest 
influence on the quality of the fits. Our calculation shows that this is
true for both the $\dres {1232}$ and the $\nres {1520}$ resonance, with the
latter showing major contributions to the $E_{0+}^{p}$ multipole. Since the
interpretation of the $z_i$ parameters from a fieldtheoretical point of view
is still unclear and since there is no prediction for the exact values there 
is clearly a need for their determination from experimental data. Here also
calculations of other reactions are neccessary because otherwise the large
number of free parameters does not allow to give strict limits on the 
extracted couplings.

\aein Furthermore we confirm the finding of Garcilazo et al. that the 
multipole data can only be reproduced by using an u-channel cutoff for the
resonance contributions. Otherwise these divergent amplitudes dominate the 
calculated multipoles for higher energies. For the different cutoffs used
in the width parametrizations one can say that the quality of the fits does
not depend one their exact form. Any cutoff that leads to a decrease of the
width beyond the resonance position would give similar $\chi^2$ values.

\aein As we have already discussed, the better fits using the resonance pole
positions show that the rescattering needs to be included in our
calculation. This would also limit the number of free parameters in the 
photoproduction since all hadronic parameters would then be uniquely defined
by the other reaction channels. 

\aein From the calculation of Sauermann et al. \cite{sauer} the influence 
of the rescattering effects in the $E_{0+}^{p}$ channel can be estimated. 
Since for most of our fits the offshell $\nres {1520}$ contributions are of 
the same order of magnitude (comp. Fig. \ref{e0pp2}) one can compare 
rescattering and offshell effects without depending on the exact choice of 
coupling parameters. In doing so one has to keep in mind that in our 
calculation part of the rescattering is already taken care of by the decay 
width of the nucleon resonances. This corresponds to 'direct' rescattering 
going through the same resonance in a K-matrix calculation. So only the 
'indirect' rescattering (through background and other resonances) is missing 
in our calculation. From \cite{sauer,sauer2} and our calculations we find 
that this 'indirect' part and the offshell $\nres {1520}$ contributions are 
of equal importance. Therefore it is not possible to neglect either of these 
effects without introducing large uncertainties in the extracted resonance 
electromagnetic couplings.

\aein Our results clearly show the need for a complete K-matrix calculation
including all resonances and multipole channels. Due to the offshell 
$\nres {1520}$ contribution the multipole decomposition does not allow
to disentangle the different nucleon resonances. Especially the extracted
$\nres {1535}$ and $\nres {1650}$ resonance parameters using the 
$S_{11}$ $\pi N$ channel and the $E_{0+}^{p}$ photoproduction multipole
\cite{sauer} are not independent of the $\nres {1520}$ couplings. Also the
high sensitivity of our fits on the resonance masses indicate that
a simultanous determination of both masses and hadronic as well as
electromagnetic couplings is needed.\\

\aein We gratefully acknowledge discussions with E. Moya de Guerra and 
H. Garcilazo on the vector meson contribution.

\newpage

%
%

\newpage

%
%

\section*{Figures}

\begin{figure}[ht]
\caption{Real and imaginary part of the $E_{0+}^{p}$ photoproduction 
multipole in millifermi units for different fits. For comparison we show 
the result of Garcilazo et al..
Solid lines: results of \protect\cite{gg}, dashed: using parameters of 
Fit 3, dashed-dotted: Fit 7. The data are taken from \protect\cite{said}.}
\label{e0pp}
\end{figure}

\begin{figure}[ht]
\caption{Same as Fig. \protect\ref{e0pp}, but for the 
$\nres{1520}$ s- and u-channel contributions only. \hspace*{1cm}}
\label{e0pp2}
\end{figure}

\begin{figure}[ht]
\caption{Electromagnetic multipoles for the isospin-$\einh$ channel and
proton target in millifermi calculated using the couplings of Fit 5 with
the resonance pole positions. The solid (dashed) lines are the real 
(imaginary) parts. Solid (open) squares are the experimental data of 
\protect\cite{said}. The marks on the energy axis indicate that for 
certain multipoles no single-energy-data was available in this energy 
range.}
\label{ipall}
\end{figure}

\begin{figure}[ht]
\caption{Same as Fig. \protect\ref{ipall} but for the neutron target.\hspace*{7cm}}
\label{inall}
\end{figure}

\begin{figure}[ht]
\caption{Same as Fig. \protect\ref{ipall} but for the isospin-$\dreih$ 
channel.\hspace*{6cm}}
\label{i32all}
\end{figure}

\newpage

\pagestyle{empty}

\setcounter{figure}{0}

\begin{figure}[htp]
\centerline{
\epsfxsize=16cm \epsfbox{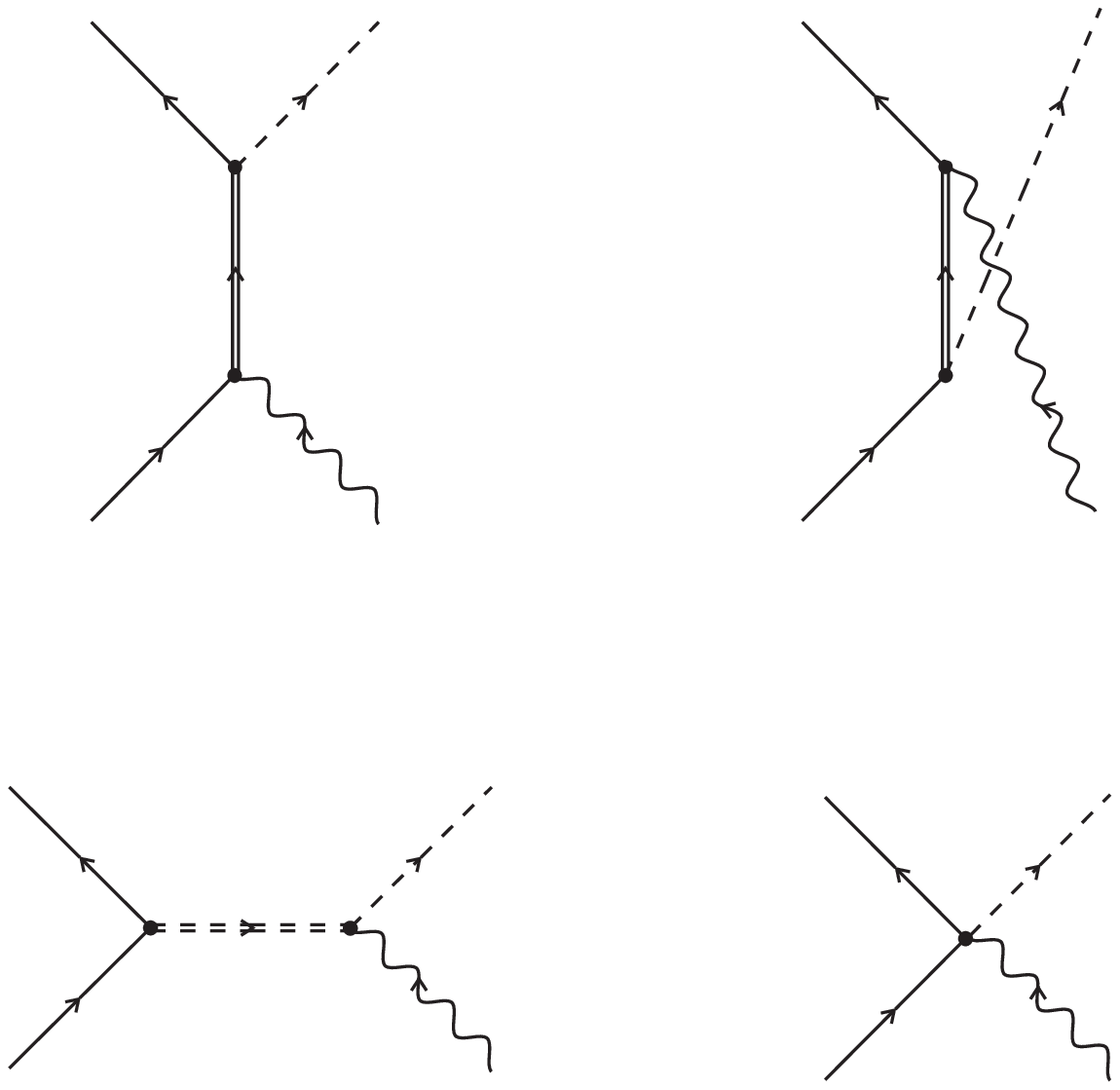}
}
\caption{Feynman diagrams for pion photoproduction. From left to right:
direct graph, exchange graph, pion pole or vector meson graph, seagull graph.}
\end{figure}

\begin{figure}[ht]
\centerline{
\epsfxsize=16cm \epsfbox{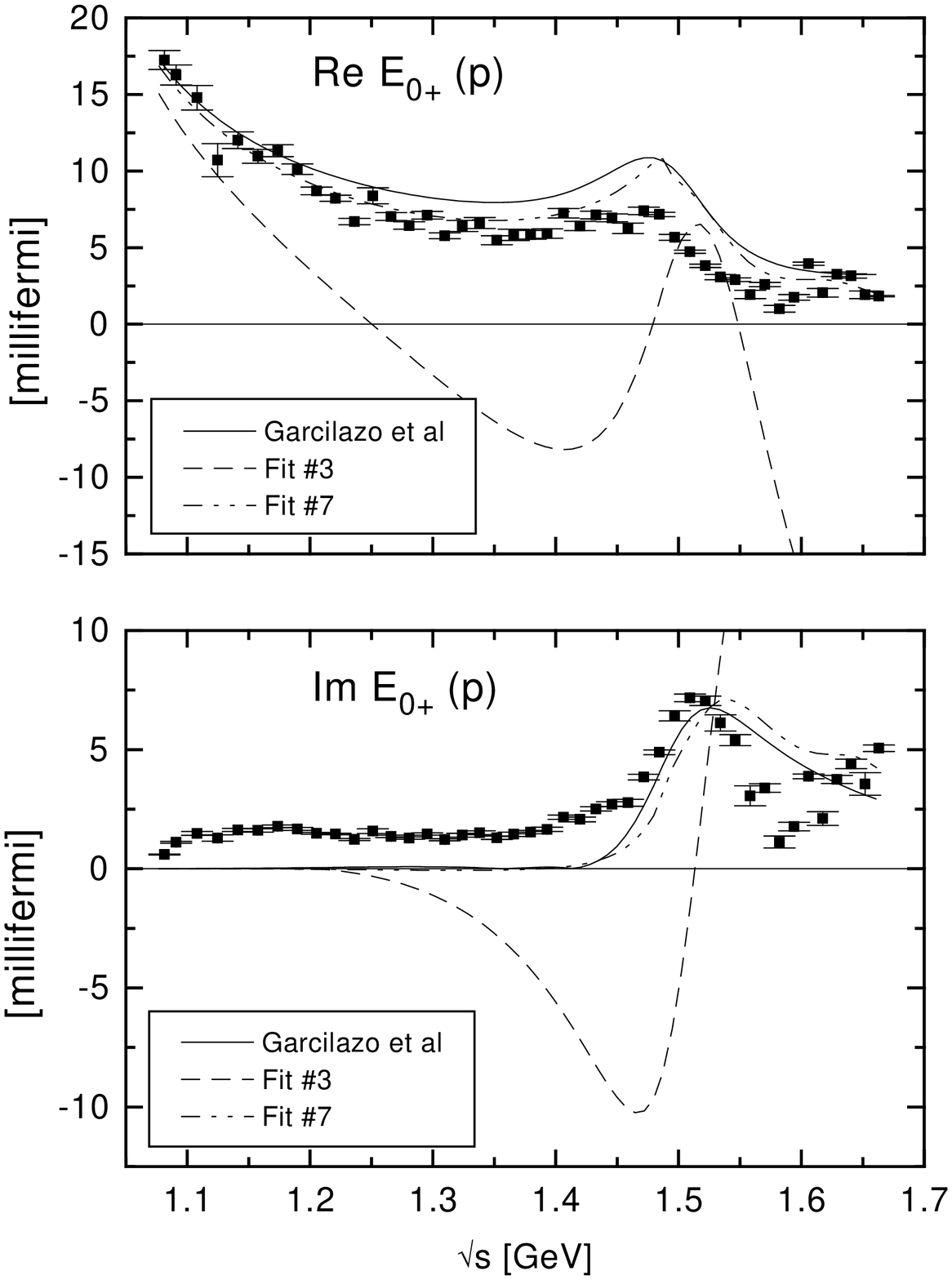}
}
\caption{$E_{0+}^p$ pion photoproduction multipole, all contributions.}
\end{figure}

\begin{figure}[ht]
\centerline{
\epsfxsize=16cm \epsfbox{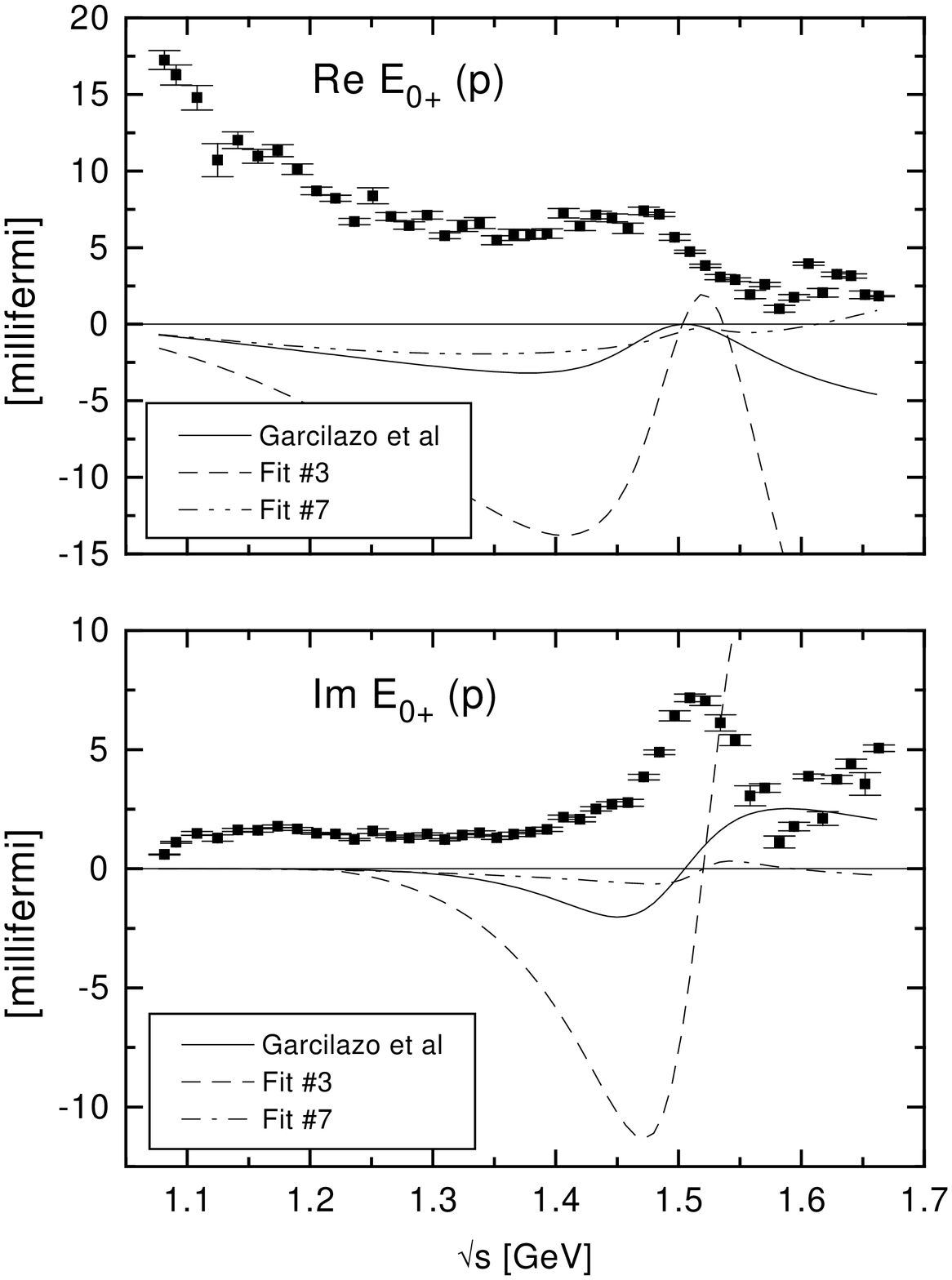}
}
\caption{$E_{0+}^p$ pion photoproduction multipole, 
$\nres{1520}$ s- and u-channel 
contributions only.}
\end{figure}

\begin{figure}[ht]
\centerline{
\epsfxsize=16cm \epsfbox{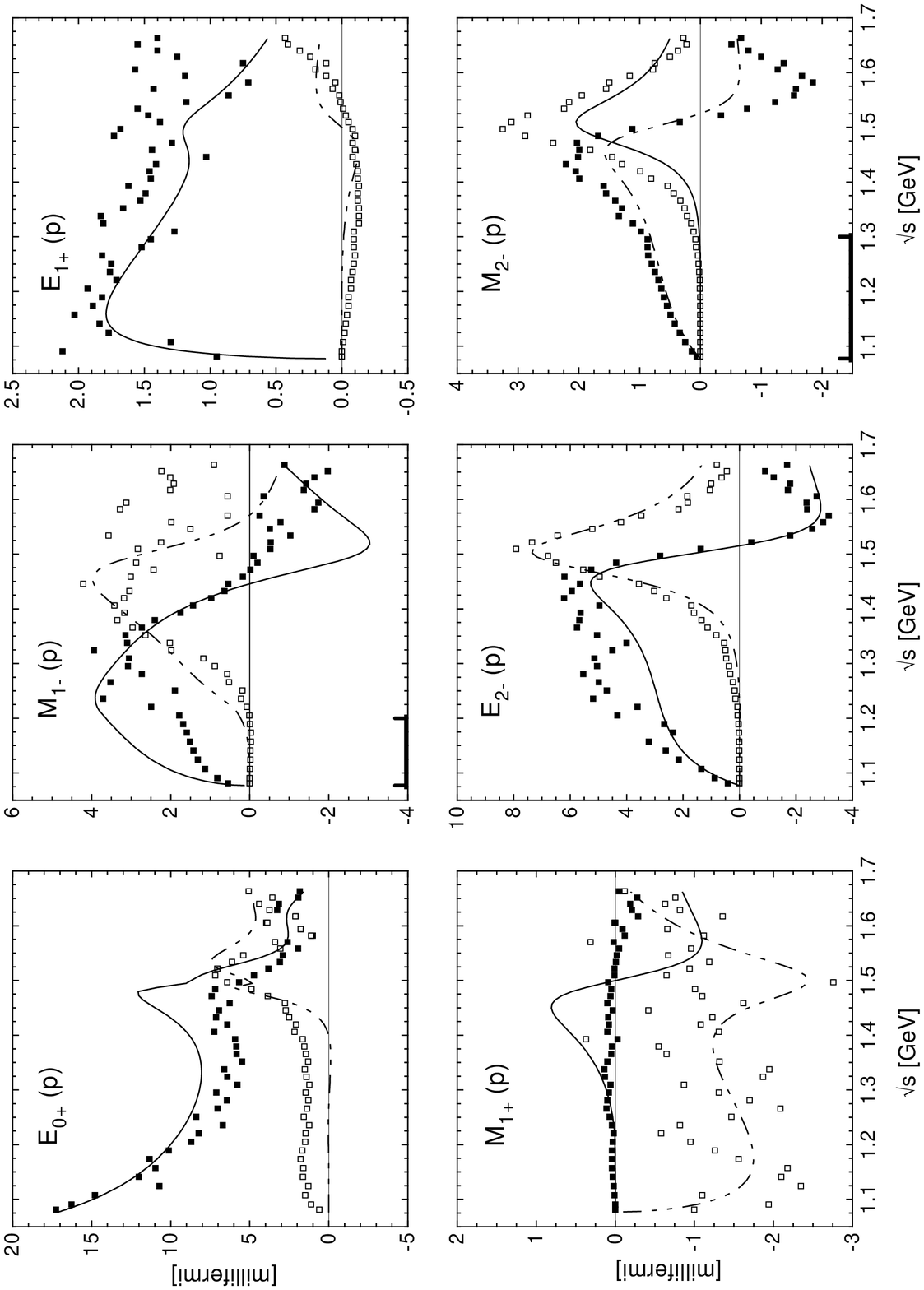}
}
\caption{Electromagnetic multipoles for the isospin-$\einh$ channel and
proton target.}
\end{figure}

\begin{figure}[ht]
\centerline{
\epsfxsize=16cm \epsfbox{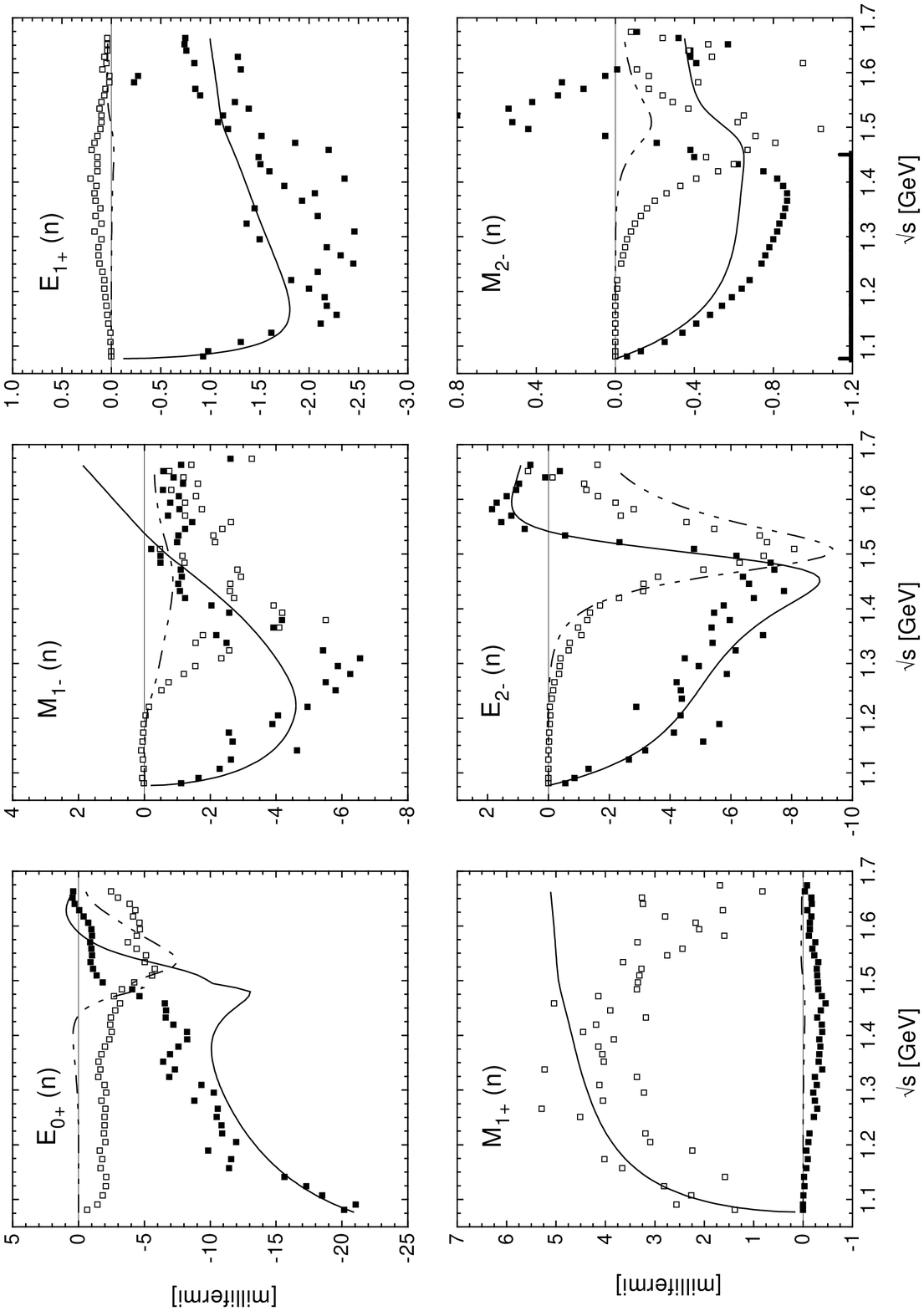}
}
\caption{Electromagnetic multipoles for the isospin-$\einh$ channel and
neutron target.}
\end{figure}

\begin{figure}[ht]
\centerline{
\epsfxsize=16cm \epsfbox{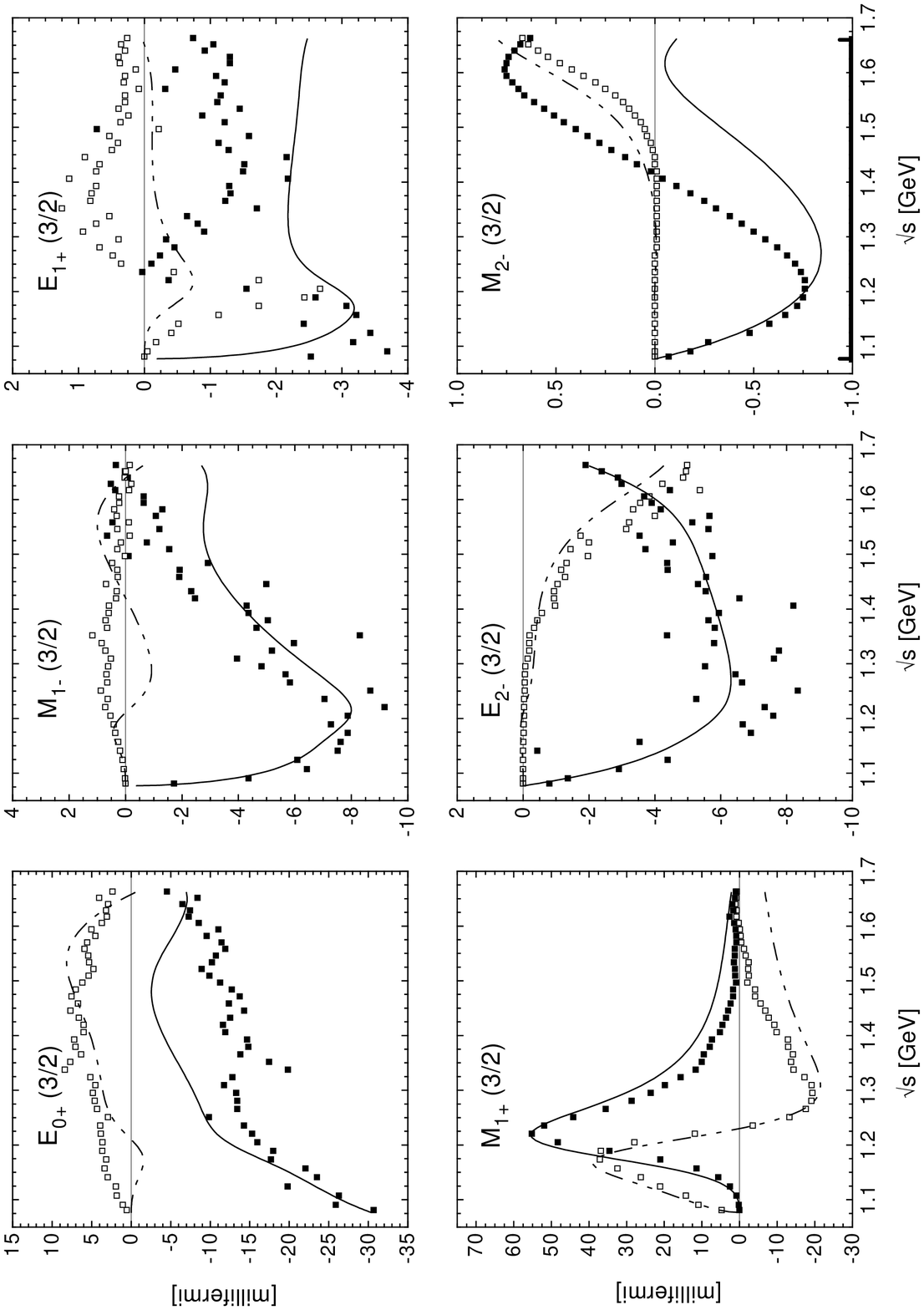}
}
\caption{Electromagnetic multipoles for the isospin-$\dreih$ channel.} 
\end{figure}

\end{document}